\documentstyle[aps,prb,amsfonts,preprint]{revtex}
\begin{document}
\title{A reliable Pad\'{e} analytical continuation method \\
based on a high accuracy symbolic computation algorithm}
\author{K. S. D. Beach\cite{byline} \and R. J. Gooding}
\address{Dept. of Physics, Queen's University, Kingston, Ontario, 
Canada K7L 3N6}
\author{F. Marsiglio}
\address{Dept. of Physics, University of Alberta, 
    Edmonton, Alberta, Canada T6G 2J1}
\date{\today}
\maketitle
\begin{abstract}

We critique a Pad\'{e} analytic continuation method whereby a rational 
polynomial function is fit to a set of input points by means of a single 
matrix inversion. This procedure is accomplished to an 
extremely high accuracy using a novel symbolic computation algorithm. 
As an example of this method in action, it is applied to the
problem of determining the spectral function of a single-particle thermal
Green's function known only at a finite number of Matsubara frequencies with
two example self energies drawn from the T-matrix theory of the Hubbard model.
We present a systematic analysis of the effects of error in the input points
on the analytic continuation, and this leads us to propose a procedure
to test quantitatively the reliability of the resulting continuation,
thus eliminating the black magic label frequently attached to this 
procedure.

\end{abstract}


\section{introduction}

Analytic continuation arises in the many-body problem whenever 
real time dynamics are to be recovered from a response function
calculated at non-zero temperatures in the Matsubara formalism.
In that case, the function whose value is known only at a discrete 
set of points on the imaginary axis must be continued to the real axis.

A general statement of the problem of interest in this paper is as follows:
An analytic continuation of a function $f$ defined on a 
subset ${\cal A} \subset {\Bbb C}$ is a function
that coincides with $f$ on ${\cal A}$ and is analytic
on a domain containing ${\cal A}$.  Usually, we are interested
in the analytic continuation $\bar{f}$ with the largest such domain,
for then $\bar{f}$ is the greatest analytic extension of $f$ 
to the complex plane.
Since there exists no general prescription for finding 
$\bar{f}$ from $f$, there is no choice but to resort to
approximate techniques.  Currently, the state of the ``art" is to 
interpolate between known points using fitting functions
capable of reproducing the analytic structure of $\bar{f}$
in the complex plane \cite {other_methods}.  A serious difficulty is that 
the analytic structure of $\bar{f}$ is not usually known a priori.

A widely used technique is the Pad\'{e} approximant method
in which ratios of polynomials (or terminating continued fractions)
are used as fitting functions.  Several Pad\'{e} schemes exist.  
The most common scheme, a recursive algorithm called
Thiele's Reciprocal Difference Method\cite{Baker},
was used by Vidberg and Serene\cite{Vidberg-Serene} 
in the context of the Eliashberg equations.
Yet, despite twenty years of widespread use,
the Pad\'{e} approximant method remains somewhat
of an untested approach in that there is still no reliable, quantitative
measure of the quality of a Pad\'{e} result.
The prevailing wisdom is that a Pad\'{e} fit
can be considered `good' when the output function is 
stable with respect to the addition of more input points.
The results of this work make it clear that such a criterion
is insufficient.

The various Pad\'{e} schemes can be divided into
two broad classes: (I) those which return the value of the 
continued function point by point in the complex plane
$(f({\cal A}),z)\mapsto \bar{f}(z)$
and (II) those which yield the function itself
$f({\cal A}) \mapsto \bar{f}$
by returning the polynomial (or continued fraction) coefficients.
Thiele's method is class I, as are most numerical methods.
In this work we present a robust Pad\'{e} scheme that is
class II and propose a goodness-of-fit criterion based on the
convergence of the polynomial coefficients to allowed values.
One advantage of our approach is that we formulate the problem
as a matrix equation, allowing us to make use of existing, highly
efficient routines for matrix inversion.  In contrast,
a na\"{\i}vely implemented recursion algorithm can lead to a severe 
propagation of error since repeated operations are performed 
on terms of very different orders of magnitude.

Our paper is organized as follows. In the next section we review
the formal aspects of thermal Green's functions to establish
the definitions of the various functions that enter into this problem.
In \S III we present the details of the Pad\'{e} form that we
will use, and state the algorithm that we use to solve for the
Pad\'{e} coefficients. This leads to the consideration of the accuracy
required for such a calculation, thus necessitating 
the use of a high accuracy symbolic computation algorithm. This
is presented in \S IV, and then we display our numerical results for
relevant test functions, including the statistical
test that allows us to conclude whether or not a given analytic
continuation is accurate. Finally, in \S V we present our conclusions.

\section{green's function formalism}

First, we introduce the components of theories based on thermal Green's 
functions to establish the definitions of the various functions that enter 
into this problem. 

The one-particle propagator or Green's function can be formulated
using real or imaginary time operators.  In real time, the retarded 
Green's function
    \begin{equation} \label{real-time-G}
        G^{\text{R}}(t) = -i\bigl\langle\bigl\{c(it),
        c^{\dagger}(0)\bigr\}\bigr\rangle\theta(t)
    \end{equation}
describes how the system responds when a particle is added
at time zero and removed at time $t$.   
Its imaginary time counterpart, the thermal Green's function        
    \begin{equation} \label{imag-time-G}
        G(\tau) = -\bigl\langle{\sf T}\bigl[c(\tau)
        c^{\dagger}(0)\bigr]\bigr\rangle\:,
    \end{equation}
is not so clearly physically motivated.  Its main advantages are
its mathematical elegance and computational ease. Further,
since it is defined in terms of the time ordering operator 
${\sf T}$, $G(\tau)$ admits a diagrammatic expansion via Wick's theorem.
Moreover, whereas the retarded Green's function $G^{\text{R}}(t)$ is 
aperiodic in $t$ (it has a lone discontinuity at $t=0$),
the temperature Green's function is periodic in $\tau$ with 
period $2\beta$.

The two Green's functions have Fourier representations: the first 
a Fourier transform
    \begin{equation} \label{fourier-retarded}
        G^{\text{R}}(t) = \frac{1}{2\pi}\int_{-\infty}^{\infty}d\omega\,
        e^{-i\omega t}G^{\text{R}}(\omega)
    \end{equation}
and the second, as a consequence of its periodicity, a Fourier series
    \begin{eqnarray} 
        G(\tau) & = &\frac{1}{\beta}\sum_{\text{odd }m}
                     e^{-im\pi\tau/\beta}G_m\\ \label{fourier-temperature} 
                & = &\frac{1}{\beta}\sum_{\omega_n}e^{-i\omega_n\tau}
                     G(\omega_n)\
    \end{eqnarray}
which, in Eq.~(\ref{fourier-temperature}), we have recast as a sum over the
Matsubara frequencies $\{\omega_n = (2n-1)\pi/\beta : n\in{\Bbb Z}\}$
of some new Fourier component $G(\omega_n)$.

The formal connection between the real and imaginary time formalisms
is the following: There exists a unique function 
$\bar{G}:{\Bbb C}\mapsto{\Bbb C}$ 
with asymptotic form 
    \begin{equation}
        \bar{G}(z) = (1/z)(1+{\cal O}(1)/\text{Im}\, z)
    \end{equation}
which takes on the values of the Fourier components of the temperature
Green's function at Matsubara points on the imaginary axis
$\bar{G}(i\omega_n) = G(\omega_n)$ and gives the Fourier transform
of the retarded Green's function just above the real axis 
$\bar{G}(\omega+i0^{+}) = G^{\text{R}}(\omega)$.
That is, Eqs.~(\ref{fourier-retarded}) and (\ref{fourier-temperature})
can be written as
    \begin{equation}
        G^{\text{R}}(t) = \frac{1}{2\pi}\int_{-\infty}^{\infty}d\omega\,
            e^{-i\omega t}\bar{G}(\omega+i0^{+})
    \end{equation}
and
    \begin{equation}
        G(\tau) = \frac{1}{\beta}\sum_{\omega_n}
        e^{-i\omega_n\tau}\bar{G}(i\omega_n)\:.
    \end{equation}
Clearly, all the information one can potentially extract from these
functions is contained in $\bar{G}$.

The function $\bar{G}$ has several interesting properties.
First, it is analytic everywhere in the complex plane 
with the exception of the real axis; this is a causality requirement. 
Second, the value of $\bar{G}$ in the upper and lower half planes 
is related by $\bar{G}(z^{*}) = \bar{G}(z)^{*}$,
which is a statement of the time reversal symmetry
between the retarded and advanced Green's functions.  Its immediate
consequence is that the imaginary part of $\bar{G}$ may be 
discontinuous across the real axis.  It also implies that
we need only know the function in either the upper or the lower half
plane since the other is a conjugated reflection of the first.
Third, $\bar{G}$ can be written as a Stieltjes/Hilbert transform
\begin{equation}
\bar{G}(z) = \int_{-\infty}^{\infty}d\omega\,\frac{A(\omega)}{z-\omega}
\end{equation}
where the spectral function, given by the magnitude of the discontinuity
in $\bar{G}$ across the real axis, viz. 
\begin{eqnarray}
A(\omega) &=& -\frac{1}{\pi}\text{Im}\, G^{\text{R}}(\omega) \nonumber \\
          &=& -\frac{1}{2\pi i}\bigl(\bar{G}(\omega+i0^{+})-
                               \bar{G}(\omega-i0^{+})\bigr)\,,
\end{eqnarray}
is non-negative and normalized to unity
\begin{equation}
A(\omega)\ge 0\,, \ \
\int_{-\infty}^{\infty}d\omega\,A(\omega) = 1\:.
\end{equation}

Typically, we are working in the Matsubara formalism and we 
calculate $G(\omega_n)$ from its self-energy (via
$G(\omega_n)^{-1} = i\omega_n-\xi-\Sigma(\omega_n)$)
which is in turn calculated from an approximate theory based on,
e.g., a diagrammatic expansion of the propagator.
From here the route to real time dynamics is somewhat circuitous:
\begin{equation}
    G(\omega_n)\stackrel{1}{\leadsto}\bar{G}(z)\stackrel{2}{\leadsto}
    G^{\text{R}}(\omega)\stackrel{3}{\leadsto} G^{\text{R}}(t)\:.
\end{equation}
(1) The first step is to analytically continue from the Fourier components
of the temperature Green's function to construct $\bar{G}$.
That this is possible, in principle, provided we know 
$G(\omega_n)=\bar{G}(i\omega_n)$
for an infinite set of points including the point at infinity,
was proved by Baym and Mermin\cite{Baym-Mermin}.  (2) Supposing that the
analytic continuation to the upper half plane can be found,
we merely evaluate it along the real axis (setting $z=\omega+i0^{+}$) 
to get $G^{\text{R}}(\omega)$.  (3) A Fourier transform then recovers 
the real time response function.

In practice, however, we do not know the values of $G(\omega_n)$ 
at an infinite number of points.  Moreover, even if we did, the theorem
of Baym and Mermin shows only the existence of a function $\bar{G}$. 
There is no general method to perform the analytic continuation --- 
hence the need for a procedure such as the Pad\'{e}.

\section{pad\'{e} approximants}

The Pad\'{e} method is based on the assumption that $\bar{G}$ can be
written as a rational polynomial or terminating continued fraction.
Since theories are most commonly specified by a choice of self-energy,
the continued fraction form turns out to be the more useful, at least
for investigating questions of a mathematical nature (e.g. analytic
structure).  In particular,
we shall find it helpful to consider $\bar{G}$ (in the upper half plane)
a continued fraction of Jacobi form\cite{Wall,usherb pades} (J-frac).  
That is,
    \begin{eqnarray} \label{J-frac}
    \bar{G}(z) = \bar{G}_{(r+1)}(z) &=& 
    \frac{\lambda_0^2}{z-e_0-}\frac{\lambda_1^2}{z-e_1-}
        \cdots\frac{\lambda_r^2}{z-e_r}\\ \label{Dyson}
    &=& \frac{1}{z-\xi-\bar{\Sigma}_{(r)}(z)}
    \end{eqnarray}
where the $\lambda_n$ and $e_n$ are complex constants.
By comparison with Dyson's equation, Eq.~(\ref{Dyson}), we make
the identification $\lambda_0^2=1$ and $e_0=\xi$, 
where $\xi$ is just the free particle energy measured with 
respect to the chemical potential\cite{Self-energy-note}. Then, we 
find that $\bar{\Sigma}_{(r)}(z)$ is itself a continued fraction
    \begin{equation} \label{Sigma-J-frac}
    \bar{\Sigma}_{(r)}(z) = \frac{\lambda_1^2}{z-e_1-}
        \frac{\lambda_2^2}{z-e_2-}
        \cdots\frac{\lambda_r^2}{z-e_r}\:.
    \end{equation}   

The justification for this continued fraction form is
a theorem due to Wall and Wetzel\cite{Wall-Wetzel} which assures us that 
a positive definite J-frac has a spectral 
representation with non-negative, integrable spectral weight
and that it is analytic in the upper half complex plane ---
all the properties we know $\bar{G}$ must have to 
be physically reasonable.  By positive definite J-frac
we mean a continued fraction in the form of Eq.~(\ref{J-frac}) 
satisfying $\text{Im}\, e_n \le 0$ and for which there exists a sequence of 
real numbers $g_0,g_1,\dots$ ($0\le g_n \le 1$) such that
\begin{equation} \label{pos-def}
(\text{Im}\, \lambda_n)^2 = (\text{Im}\, e_{n-1})(\text{Im}\, e_n)(1-g_{n-1})
g_n\:.
\end{equation}

There are two special cases worth mentioning.
If the $\lambda_n$ and $e_n$ are all real then the J-frac is 
positive definite and can be cast 
as a sum of simple poles\cite{Wall}
       \begin{equation}
           \sum_{n=1}^{r}\frac{R_n}{z-E_n}
       \end{equation}
with real, distinct energies $E_n$ and positive residues $R_n>0$.
The J-frac is also positive definite if the $\lambda_n$ are real 
and none of the $e_n$ sits in the upper half complex plane 
(Eq.~(\ref{pos-def}) is satisfied by setting all $g_n = 0 \ \text{or} \ 1$),
in which case the function is characterized by simple poles resting
on or below the real axis.
In the general case, all the continued fraction coefficients have the
potential to be complex, with the exception of $\lambda_0^2=1$,
$e_0=\xi$, and $\lambda_1^2$.  Since $e_0 = \xi$ has no imaginary 
part, Eq.~(\ref{pos-def}) implies that the coefficient $\lambda_1^2$ 
must always be real and positive.

It is clear that by observing the values of the $\lambda_n$, $e_n$
coefficients, one can learn a great deal about the analytic properties 
of $\bar{G}_{(r+1)}$.  For example, if some $e_n$ has a positive imaginary
part (and no $\lambda_m=0$ for $m<n$) then $\bar{G}_{(r+1)}$ may have a
pole in the upper half plane --- such a function would be noncausal and
have negative spectral weight. (In fact, it is through such considerations 
that we are led to propose a method for testing the accuracy of a given
analytic continuation via a Pad\'{e}.)

Nonetheless, despite the usefulness of the continued fraction form,
for computational purposes it is actually much easier to work with
rational polynomials.  Conveniently, every terminating continued 
fraction is equivalent to a rational polynomial.  For instance, a 
J-frac with $r$ stories, Eq.~(\ref{Sigma-J-frac}) say, can be written 
as the ratio
    \begin{equation} \label{rational-poly}
        \bar{\Sigma}_{(r)}(z) = \frac{P_{(r)}(z)}{Q_{(r)}(z)}
    \end{equation}
    of two polynomials $P,Q$ defined recursively
by the formulas
    \begin{mathletters} \label{inductive}
    \begin{equation}
        P_{(n+1)} = (z-e_n)P_{(n)}(z)-\lambda^2_nP_{(n-1)}(z)
    \end{equation}
    \begin{equation}
        Q_{(n+1)} = (z-e_n)Q_{(n)}(z)-\lambda^2_nQ_{(n-1)}(z)
    \end{equation}
    \end{mathletters}
(for $n=1,2,3,\ldots$) with base cases
    \begin{mathletters} \label{base}
    \begin{equation}
        P_{(0)}=0, \: P_{(1)} = \lambda_1^2
    \end{equation}
    \begin{equation}
        Q_{(0)}=1, \: Q_{(1)} = z-e_1\:.
    \end{equation}
    \end{mathletters}
Writing out the leading order terms of $P$ and $Q$
    \begin{mathletters}
    \begin{equation}
        P_{(r)}(z) = \lambda_1^2z^{r-1} - 
        \lambda_1^2(e_2+e_3+\cdots+e_r)z^{r-2} + \cdots
    \end{equation}
    \begin{equation}
        Q_{(r)}(z) = z^r - (e_1+e_2+\cdots+e_r)z^{r-1} + \cdots
    \end{equation}
    \end{mathletters}
makes it clear that the polynomial $P$ is of order $r-1$ in $z$
while the polynomial $Q$ is of order $r$.  (Accordingly, one refers
to $\bar{\Sigma}_{(r)}$ in Eq.~(\ref{rational-poly}) as a
$[r-1/r]$ rational polynomial.)  Moreover, it suggests that we 
write the self energy explicitly as a rational polynomial of the form
    \begin{equation}
        \bar{\Sigma}_{(r)}(z) = \frac{p_1+p_2z+\cdots+p_rz^{r-1}}
        {q_1+q_2z+\cdots+q_rz^{r-1}+z^r}\:.
    \end{equation}
It is straightforward to relate the 
old and new coefficients to one another via Eqs.~(\ref{inductive}) 
and (\ref{base}): e.g. $\lambda_1^2 = p_r$, $e_1 = p_{r-1}/p_r-q_r$, etc.

The coefficients $p_n$, $q_n$ can be determined by specifying the value 
of $\bar{\Sigma}_{(r)}$ at $2r$ points, {\it viz.},  by solving the set 
of $2r$ linear equations\cite{Linear-note}
    \begin{equation} \label{sys-eqs}
        \bigl\{\bar{\Sigma}_{(r)}(i\omega_n) = \Sigma(\omega_n)\bigr\}\:.
    \end{equation}
If we define the column vectors
    \begin{equation}
        \left[ \begin{array}{c} {\bf p}\\ {\bf q} \end{array}\right]
        = \left[\begin{array}{c} p_1 \\ \vdots \\ p_r \\ q_1 \\ \vdots \\ 
        q_r \end{array}\right]
        \hspace{0.1in}\text{and}\hspace{0.1in}
\tilde{{\boldmath \mbox{$\sigma$}}} = \left[\begin{array}{c} 
\sigma_1(i\omega_1)^r \\ 
\sigma_2(i\omega_2)^r\\ \vdots \\ 
\sigma_{2r}(i\omega_{2r})^r \end{array}\right]\:,
    \end{equation}
where $\sigma_n = \Sigma(\omega_n)$ are the known values of the self-energy
at $2r$ Matsubara frequencies, and a matrix
    \begin{equation} \label{Xmatrix}
        X = \left[\begin{array}{ccccccc}
1 & i\omega_1 & \cdots & (i\omega_1)^{r-1} & -\sigma_1 & \cdots & 
-\sigma_1(i\omega_1)^{r-1}\\
1 & i\omega_2 & \cdots & (i\omega_2)^{r-1} & -\sigma_2 & \cdots & 
-\sigma_2(i\omega_2)^{r-1}\\
\vdots & & & & & & \vdots\\
1 & i\omega_{2r} & \cdots & (i\omega_{2r})^{r-1} & -\sigma_{2r} & 
\cdots & -\sigma_{2r}(i\omega_{2r})^{r-1}
        \end{array}\right]%
    \end{equation}
equivalent to the system of equations given by Eq.~(\ref{sys-eqs}),
then the entire process of analytic
continuation is reduced to a single matrix inversion
    \begin{equation}
        \left[\begin{array}{c} {\bf p}\\ {\bf q} \end{array}\right] 
         = X^{-1}\tilde{\boldmath \mbox{$\sigma$}}
    \end{equation}
which provides the polynomial coefficients necessary to construct
    \begin{equation}
  \bar{\Sigma}_{(r)}(z) = \frac{\left[\begin{array}{ccccc} 1 & z & z^2 & 
  \cdots & z^{r-1}\end{array}\right]{\bf p}}
{\left[\begin{array}{ccccccc} 1 & z & z^2 & \cdots & z^{r-1}\end{array}\right]
  {\bf q}+z^r}\:.
    \end{equation}

What we propose is that, having determined the $p_n$, $q_n$ coefficients, 
we recover the $\lambda_n$, $e_n$ coefficients and 
then use the criteria provided by Wall and Wetzel's theorem 
to determine whether the matrix inversion produced a $\bar{G}_{(r+1)}$
with an acceptable analytic form.  As a first step, we investigate what
can be learned from $\lambda_1^2$, the first non-trivial J-frac coefficient.
$\lambda_1^2$ is equal to the sum of the residues of the poles in the
self-energy and as such it gives the high frequency asymptotic
behaviour of the self-energy via $\bar{\Sigma}_{(r)}(z) \sim 
\lambda_1^2/z$.  A necessary condition for positive definiteness is
that $\lambda_1^2$ be real and positive.  
We shall see that the convergence of $\text{Im}\, \lambda_1^2$ to zero 
as a function of the number $r$ of poles in the Pad\'{e} fitting function
can provide information on the quality of the fit and on the analytic
structure of the true continuation $\bar{G}$.

\section{numerical results}

The procedure we have outlined in Sect.~III is a specialization of the
following general Pad\'{e} procedure --- such considerations are
central to our statistical analysis of the quality of the fits
provided by this method.

Given a function $f$ and a set ${\cal A}$ of $2r$ input points, 
we suppose that we can
approximate the analytic continuation $\bar{f}$ by a $[r-1/r]$
rational polynomial $\bar{f}_{(r)}$, the coefficients of which
are determined by solving the linear system of equations
$\{\bar{f}_{(r)}(a)=f(a):a\in{\cal A}\}$.
This problem can be cast as a matrix inversion in which the kernel
$X$ has elements with ratios as large as
\begin{equation}
\zeta = |(\max {\cal A}\cup f({\cal A}))^{r-1}/
\min {\cal A}\cup f({\cal A})|\:.
\end{equation}
Thus to reliably perform
the inversion we need a numerical range $\sim \zeta^2$,
i.e. $2\log_{10}\zeta$ decimal digits of numerical precision.
This analysis is general in that no other Pad\'{e} algorithm
can have less stringent precision requirements.

For the case of a self-energy $\Sigma$, 
known at the first $2r$ Matsubara frequencies 
above the real line on the imaginary axis, 
we have shown that the matrix $X$ is given by Eq.~(\ref{Xmatrix}).
Since $\Sigma(\omega_n)\sim 1/\omega_n$, the
ratio of the largest to smallest terms in $X$ is 
$\zeta = (\omega_{2r})^r=((4r-1)\pi T)^r$, the square of
which gives an estimate of the amount of 
precision needed to invert $X$.  
Here, that corresponds to 
    \begin{equation} \label{dig-precision}
        2r \log_{10} (4r-1)\pi T
    \end{equation}
decimal digits.

To achieve a sufficient level of precision for our numerical work, we
implement the Pad\'{e} algorithm using the symbolic computation package
MAPLE.  Under MAPLE, expression evaluation takes place in software and thereby
transcends the limits imposed by hardware floating-point.  All computations
are performed in base ten to any desired level of precision
(we specified {\tt Digits~:=~250;} \cite {250}).
Moreover, MAPLE is an ideal environment for rapid prototyping since high 
level matrix data types and routines are available as primitives.

We begin by considering a test function of known analytic 
structure.  The self-energy\cite{Mat-freq-note}
    \begin{equation} \label{T-matrix-sigma}
    \Sigma(\vec{k},\omega_n) = -\frac{U^2}{\beta
        M}\sum_{\vec{Q}}\sum_{\nu_{n'}}
        \chi^0(\vec{Q},\nu_{n'})G^0(\vec{Q}-\vec{k},\nu_{n'}-\omega_n)
    \end{equation}
corresponds to the first `rung' of the ladder diagrams in the 
T-matrix \cite{T-matrix-ref} approximation of the single-band Hubbard 
model\cite{hubbard} (characterized by a near-neighbour hopping integral $t$
and an on-site repulsion energy $U$). Here, $G^0$ is the free propagator
    \begin{equation}
        G^0(\vec{k},\omega_n) = \frac{1}{i\omega_n - \xi_{\vec{k}}}
    \end{equation}
and $\chi^0$ is the free pair susceptibility
    \begin{equation} \label{pair-suscept}
\chi^0(\vec{Q},\nu_n) = \frac{1}{\beta M}\sum_{\vec{k}}\sum_{\omega_{n'}}
G^0(\vec{k},\omega_{n'})G^0(\vec{Q}-\vec{k},\nu_n-\omega_{n'})\:.
    \end{equation} 
The frequency sums in Eqs.~(\ref{T-matrix-sigma}) and (\ref{pair-suscept}) can 
be performed analytically, giving \cite {identity}
\begin{equation}
\chi^0(\vec{Q},\nu_n) = \frac{1}{M}\sum_{\vec{k}}\frac{f[\xi_{\vec{k}}]
+f[\xi_{\vec{Q}-\vec{k}}]-1}{i\nu_n - \xi_{\vec{k}} - \xi_{\vec{Q}-\vec{k}}}
\end{equation}
and
\begin{equation} \label{Sigma-exact}
\Sigma(\vec{k},\omega_n) = \frac{U^2}{M^2}\sum_{\vec{Q},\vec{k}'}
\frac{(f[\xi_{\vec{k}'}]+f[\xi_{\vec{Q}-\vec{k}'}]-1)
f[\xi_{\vec{Q}-\vec{k}}]-
f[\xi_{\vec{k}'}]f[\xi_{\vec{Q}-\vec{k}'}]}
{i\omega_n + \xi_{\vec{Q}-\vec{k}} - \xi_{\vec{k}'} 
- \xi_{\vec{Q}-\vec{k}'}}\:.
\end{equation}
Since the $\xi_{\vec{k}}$ are real, the analytic continuation of
the self-energy is a meromorphic function with a finite number of simple poles, 
all situated along the real axis.  Calculated in two dimensions on an 
$8\times 8$ ($M=64$) lattice, its $\vec{k}=0$ component 
possesses $r_0=26$ poles. 
(The number of poles is determined by counting the number of distinct
elements in the set $\{\xi_{\vec{Q}}-\xi_{\vec{k}'}-\xi_{\vec{Q}-\vec{k}'}
: \forall\,\vec{k}',\vec{Q}\}$.) 

For a particular set of parameters\cite{Param-choice-note} --- we use an
interaction strength $|U|/t=4$, chemical potential $\mu/t = -2$, and 
temperature $T/t=0.7$ 
--- the test function, Eq.~(\ref{T-matrix-sigma}), is calculated in two 
different ways for the Matsubara frequencies 
$\{\omega_1,\omega_2,\ldots,\omega_{2r}\}$.  
First, it is calculated exactly, as prescribed
by Eq.~(\ref{Sigma-exact}), but with a small, random error, {\it viz.},
each value is multiplied by $1+\epsilon$ with $-1 \leq \epsilon \leq 1$. 
Second, it is calculated by truncating the Matsubara sum at an arbitrary
cutoff frequency $\nu_p \gg 1$ (much larger than the relevant
energy scale of the problem) and then systematically 
adding back the high frequency contributions up to
a given order.  That is,
\begin{eqnarray}\nonumber
-\frac{1}{\beta}\sum_{\nu_{n'}}\chi^0(\vec{Q},\nu_{n'})
G^0(\vec{Q}-\vec{k},\nu_{n'}-\omega_n)
& = & -\frac{1}{\beta}\sum_{|\nu_{n'}|\le \nu_p}
\chi^0(\vec{Q},\nu_{n'})G^0(\vec{Q}-\vec{k},\nu_{n'}-\omega_n)\\ 
\label{my-method}
& & +\sum_{l=1}^{m-1}\chi^0_{(l)}(\vec{Q})\Theta_{(l+1)}[i\omega_n
+\xi_{\vec{Q}-\vec{k}}] + {\cal O}(1/(\nu_p)^m)
\end{eqnarray}
where
\begin{equation}
\chi^0_{(l)}(\vec{Q}) = \frac{1}{M}\sum_{\vec{k}}(f[\xi_{\vec{k}}]
+f[\xi_{\vec{Q}-\vec{k}}]-1) (\xi_{\vec{k}}+\xi_{\vec{Q}-\vec{k}})^{l-1}
\end{equation}
are the coefficients of a Laurent expansion of $\chi^0(\vec{Q},\nu_n)$ and the
$\Theta_{(l)}$ functions (defined in the Appendix) are constructed using the 
symbolic manipulation capabilities of MAPLE.

Now, we let the self-energy, evaluated at the first $2r$ Matsubara frequencies
according to the two schemes described above, 
serve as the input to the Pad\'{e} procedure. The resulting approximant
$\bar{\Sigma}_{(r)}$ yields a propagator 
$\bar{G}_{(r+1)}(\vec{k},z)=(z-\xi_{\vec{k}}-
\bar{\Sigma}_{(r)}(\vec{k},z))^{-1}$ 
with spectral function
$A_{(r+1)}(\vec{k},\omega) = -(1/\pi)\text{Im}\, 
\bar{G}_{(r+1)}(\vec{k},\omega+i0^+)$.
The spectral function derived from the Pad\'{e} approximant is compared to 
that of the exact function using the logarithmic measure
\begin{eqnarray}
10^{-F}
       &\equiv& \int_{-\infty}^{\infty}dx\,(A(\vec{k},x)
-A_{(r+1)}(\vec{k},x))^2\nonumber \\
       &=& \frac{1}{\pi^2}\int_{-\infty}^{\infty}dx\,
\Bigl|\text{Im}\, \Bigl(\bar{G}(\vec{k},x+i\eta)-
            \bar{G}_{(r+1)}(\vec{k},x+i\eta)\Bigr)\Bigr|^2 \:.
\label{eq:F eqn}
\end{eqnarray}
In practice we choose $\eta$ to be a small, but noninfinitesimal positive
real quantity (we use $\eta/t = 0.064$), 
which has the effect of introducing a slight 
artificial broadening to the $\delta$-function peaks of the spectral function.

The results of this comparison (for the $\vec{k}=0$ component of the
spectral function) are presented in Figs.~1(a) and 1(b), where $F$ is 
plotted as a function of $r$ for different values of the random error 
$E = -\log_{10}\epsilon$ 
and the systematic error $E= - \log_{10} 1/(\nu_p)^m = m \log_{10} \nu_p$
(and thus a larger $E$ corresponds to a smaller error).  
In each graph, a vertical dashed line marks the exact
number of poles ($r_0 = 26$) in the true self-energy.
The most distinctive feature of both graphs is that, at high
accuracy (large $E$), the $F$ curves exhibit a large step
at the point $r=r_0$.  In the random error case, the $E=120$ curve
jumps by four decades, and this represents an improvement in the 
Pad\'{e} fit of nearly 40 orders of magnitude.  In the systematic
error case, the result is even more dramatic: 
the $E=100$ and $E=120$ curves jump by roughly four and seven decades,
respectively.

At these large accuracies, the only factor inhibiting the success of the
Pad\'{e} approximants is the lack of a sufficient number of poles to 
reproduce the analytic structure of the true function. The large jump
observed in the large $E$ curves marks the point, $r=r_0$, at which the 
number of poles in the Pad\'{e} approximant exactly matches the required 
number, and for this and larger $r$ there is no difficulty in finding
an excellent fit of the test function. In contrast, when the input points 
are known to relatively low accuracy, no such feature is observed, and 
instead the $F$ curves pass smoothly through $r_0$. This makes clear
that for self-energies calculated to 20, 40, or even 60 decimal digits of 
accuracy, the level of error in the input points is still the main obstacle 
to a successful Pad\'{e} fit.

The usual response to this situation is to increase the number of
Pad\'{e} points in an attempt to overcome the intrinsic error limitations
(by making the system of equations more and more overcomplete).
However, whatever advantage this additional information brings to the
Pad\'{e} approximant is soon outweighed by the accompanying complications:
When a rational polynomial of degree $[r-1/r]$ is used
to fit a function with $r_0 < r$ poles,
$r-r_0$ zeros of the numerator must coincide with
an equal number of zeros in the denominator in order
to cancel the extraneous poles.  As $r-r_0$ grows,
it is less and less likely that this cancellation will
be complete. A slight misplacement of zeros leads to `defects' in which
the function moves between $0$ and $\infty$ in a small
neighbourhood.  Moreover, it cannot be predicted where these zero-zero pairs 
will appear\cite{defect-note}. For the purposes of calculating a spectral 
function, they are of little consequence provided that they lie deep in the 
complex plane.
However, when they are not so far removed from the real axis, they can distort 
the spectral function away from its proper shape.  When they lie on or near 
the real axis, they can give rise to deep troughs of negative spectral weight 
and other spurious, non-physical features.

The deterioration of the Pad\'{e} fit, as described above, 
is evident in Figs.~1(a) and 1(b) in which many of the $F$ curves
reach maxima at points $r_{\text{best}} > r_0$ and then quickly begin 
to fall off for larger $r$.  Interestingly, this behaviour is much more 
pronounced in the systematic error case where such maxima occur
for each curve.  In the random error case, the curves below
some error threshold are essentially flat for all $r$.  

The primary lesson that one should draw from these results
is that the addition of Pad\'{e} points well beyond the required
number is not a useful strategy for improving the Pad\'{e} fit.
Unless the exact analytic continuation is already known,
there is no way to predict the value of $r_{\text{best}}$.
We believe that better results are achieved by fixing the number of 
Pad\'{e} points at $2r_0$ (giving rise to a $[r_0-1/r_0]$ rational
polynomial) and working towards increasing the accuracy with which those 
input points are calculated.  Even a small effort there
can result in an improvement of several orders of magnitude in the fit.
What to try when one does not know {\it a~priori} what $r_0$ is
discussed later in this paper.

Now consider Figs.~2(a) through 2(d) in which the spectral function of a 
Pad\'{e} approximant with 26 poles (calculated by specifying the value 
of the self-energy at 52 Matsubara frequencies) is compared to the exact 
spectral function.  In Fig.~2(a), the accuracy of the input points is given by 
$E=16$ (random error), roughly the number of digits in a double precision 
Fortran variable. Despite the fact that the overall energy scale is correct,
the details of the fit are quite poor.  Here, the effect of 
insufficient accuracy is to produce a washed out version of the spectral 
function which completely lacks fine structure.
Even at $E=30$ (Fig.~2(b)), corresponding to the number of digits available
in the largest Fortran data type, the Pad\'{e} inversion is only 
just beginning to distinguish the main peaks of the spectral function.
Figure~2(c) shows the result for $E=80$ and Fig.~2(d) the result
for $E=120$.  Notice that in Fig.~2(d), the fit is near perfect: 
even the smallest peaks have been reproduced faithfully. 

In this example, with $r=r_0$, the Pad\'{e} approximant provides a remarkable
fit to the true function whenever the accuracy of the input points
is better than $E \sim 110$.  The difficulty in translating our success
in this specific case to the general problem is that, in real applications,
one has no way to judge when sufficient accuracy has been achieved.
Also, in most instances, the number of poles in the self-energy
is unknown.

In what follows, we hope to address these deficiencies.  We begin by
defining a logarithmic measure of the imaginary part of the J-frac 
coefficient $\lambda_1^2$:
\begin{equation}
10^{-\Lambda} \equiv \bigl| \text{Im}\,\lambda_1^2 \bigr| \:.
\end{equation}
We argued in Sect.~III that $\lambda_1^2$ ought to be real and positive.
In a Pad\'{e} calculation, however,  it is real-valued only to within 
some small fraction which characterizes the numerical sensitivity
of the matrix inversion.  As we shall soon discover, the convergence
of the imaginary part of $\lambda_1^2$ to zero ($\Lambda \rightarrow \infty$)
can be used (1) to determine when the threshold of accuracy for an
exact fit has been reached and (2) to infer the value of $r_0$
if it is unknown.

In Figs.~3(a) and 3(b), we plot $\Lambda$ as a function of $r$ for the 
random and systematic error cases.  Over each plot is superimposed a
reference line given by Eq.~(\ref{dig-precision}).  What we observe is
a set of $\Lambda$ curves that initially follow the reference line
but later fan out, spaced according to their $E$ values.  Our claim
is that these curves provide the quantitative measure of success of 
the Pad\'{e} approximant that has heretofore
been lacking, the essential point being that the shape of the curves
reveals the performance characteristics of the Pad\'{e} inversion
in the various $r$ regimes.

When $0<r<r_0$, the accuracy of the Pad\'{e} approximant is matrix inversion
dominated and the
behaviour of $\Lambda$ is governed by $\Lambda \sim 2r\log_{10} (4r-1)\pi T$.
In this regime, the Pad\'{e} approximant has too few poles to fit the 
true function and thus the matrix inversion must judiciously arrange the
available poles (sometimes apportioning one pole to a region where
there should be two or three) to give the best possible fit.
In the opposite limit, $r \gg r_0$, the accuracy of the Pad\'{e} approximant
is input-point error dominated.
In this regime, there are more than enough poles to perform an exact fit,
but the proper placement of those poles and the determination of their
residues is hampered by the finite accuracy to which the input points
are known.  We find this reflected in the $\Lambda$ curves which, for
large $r$, saturate at a value $\Lambda \sim E$ (roughly).

Most interesting, though, is the behaviour of $\Lambda$
in the vicinity of $r=r_0$ where the $\Lambda$ curves in 
Figs.~3(a) and 3(b) first cross the reference line.  In those plots,
we see that the $\Lambda$ curves corresponding to small values of $E$
closely follow the reference line (Eq.~(\ref{dig-precision})) 
until finite accuracy becomes a limiting 
factor.  The curves then fall below the reference line and become 
more or less flat.
As $E$ is increased, the $r$ coordinate at which a given $\Lambda$ curve 
first deviates from the reference line moves to the right
until (for some accuracy, $E_0$ say) it coincides with $r_0$.
Here, there is a sudden change in behaviour: all $\Lambda$ curves
corresponding to accuracies $E>E_0$ cross the reference line
at $r=r_0$.  Such a crossing signals that there are now both 
sufficient poles in the approximant and sufficient accuracy on the 
input points to fit $\bar{\Sigma}$ more or less exactly.  We can verify this 
interpretation by appealing to Figs.~1 and 2 which clearly show a large 
jump at $r_0$ for precisely the same curves that demonstrate a crossing 
in Figs.~3(a) and 3(b).

The results we have described are extremely general and do
not dependent on the choice of test function.  For example,
we may replace Eq.~(\ref{T-matrix-sigma}) with the full non-self-consistent
T-matrix self-energy
    \begin{equation}\label{full-NSC-T-matrix-self-energy}
    \Sigma(\vec{k},\omega_n) = -\frac{U^2}{\beta
        M}\sum_{\vec{Q}}\sum_{\nu_{n'}}
        \frac{\chi^0(\vec{Q},\nu_{n'})G^0(\vec{Q}-\vec{k},\nu_{n'}-\omega_n)}
        {1+U\chi^0(\vec{Q},\nu_{n'})}\:.
    \end{equation}
Here, the frequency sums cannot be performed analytically\cite{Freq-sum-note}
and thus we do not have a closed form analytical expression for the 
self-energy.
(Thus, this is more representative of the usual situation in which the Pad\'{e} 
method might be applied.)  In this case, we know only that its analytic
continuation has a finite number of poles along the real axis (although we
are able to predict analytically an upper bound for the number of poles).

This self-energy can be calculated to high accuracy 
using the method of Eq.~(\ref{my-method}) with
the $\chi^0_{(l)}(\vec{Q})$ replaced by the 
coefficients of the Laurent expansion of 
$\chi^0(\vec{Q},\nu_n)/(1+U\chi^0(\vec{Q},\nu_n))$.
That is, $\chi^0_{(1)}(\vec{Q}) \mapsto \chi^0_{(1)}(\vec{Q})$,
$\chi^0_{(2)}(\vec{Q}) \mapsto \chi^0_{(2)}(\vec{Q})-U\chi^0_{(1)}(\vec{Q})^2$
and so on according to \cite{series}
\begin{equation} 
\frac{\frac{\chi_{(1)}}{i\nu_n}+\frac{\chi_{(2)}}{(i\nu_n)^2}
+\frac{\chi_{(3)}}{(i\nu_n)^3}+\cdots}
{1+U\Bigl( \frac{\chi_{(1)}}{i\nu_n}+\frac{\chi_{(2)}}{(i\nu_n)^2}
+\frac{\chi_{(3)}}{(i\nu_n)^3}+\cdots \Bigr)}
= \frac{\chi_{(1)}}{i\nu_n}+\frac{\chi_{(2)}-U\chi_{(1)}^2}{(i\nu_n)^2}
+\frac{\chi_{(3)}-2U\chi_{(2)}\chi_{(1)}+U^2\chi_{(1)}^3}{(i\nu_n)^3}+\cdots
\end{equation}

The Pad\'{e} approximant method can then be applied to
Eq.~(\ref{full-NSC-T-matrix-self-energy}) calculated in this way.
We find that the resulting plot of $\Lambda$ vs.\ $r$ is identical 
to that of Fig.~3(b) except that the crossing of the reference line
at high accuracy now occurs at $r=156$.
This allows us to deduce that the function has $r_0=156$ poles, 
significantly more than the 26 poles of Eq.~(\ref{Sigma-exact}).
(This is a consequence of the lifting of degeneracy in each $\vec{Q}$ 
component brought about by the renormalization $1/(1+U(\vec{Q},\nu_n))$.)
We also find that the approximant spectral function compares well
with increasing accuracy of the input points to the numerically exact 
spectral functions as 
calculated (i) by a non-Pad\'{e} method due to 
Marsiglio et al.~\cite{Marsiglio}
(this non-Pad\'{e} method is of limited application since it requires 
the self-energy to have a very specific form, but for those cases 
where it is applicable, it can outperform the Pad\'{e} method),
and (ii) by an exact partial fraction decomposition of the self energy 
\cite {sc paper} that can be done to a very high accuracy 
(say $10^{-40}$ on all poles and residues).

Finally, one interesting feature that could potentially be exploited
is that for self-energy values calculated using the $\Theta$ function
expansion, the value of $r$ which gives the maximum value of $\Lambda$
roughly tracks $r_{\text{best}}$ (cf.\ Figs.~1(b) and 3(b)).

\section{conclusions}

The Pad\'{e} procedure is very sensitive to the numerical precision
with which the matrix inversion is performed and to the
intrinsic error on the input points.  Sufficient precision is difficult 
to achieve in traditional computer languages (e.g.\ C, Fortran) and so, 
in many instances, it may be necessary to make use of a symbolic computation
package capable of supporting very large precision data types.
Likewise, sufficient accuracy is difficult to achieve without a sophisticated
computational scheme (e.g.\ the $\Theta$ function expansion)
that goes beyond a simple truncation of the Matsubara frequency sums in
the self-energy.  The required level of precision and accuracy depends
on the temperature $T$, which controls the spacing of the Matsubara
points, and on the pole count $r_0$.

An insufficient level of accuracy leads to an approximant spectral function 
that lacks fine structural detail or, worse, one that exhibits spurious 
spikes or troughs of spectral weight.
This poses a problem whenever we are interested 
in the presence of a specific feature in the spectral function (e.g.\ 
the onset of a normal state pseudogap).  In that case,
it is essential to have confidence in the quality of the Pad\'{e} result.
We must be convinced that the observed feature is robust and 
not merely a by-product of insufficient accuracy.

We have argued that simply adding more Pad\'{e} points cannot compensate 
for too large an error on the input points.  While there is a small set of
$r$ values for which an increase in $r$ improves the fit, there is no known
criterion that indicates when to stop adding points.  
Without already knowing the exact result, 
one cannot distinguish between the regime where additional points improve 
the fit ($r < r_{\text{best}}$) and the regime where such points degrade it 
($r\ge r_{\text{best}}$).  Instead, we recommend
the use of a Pad\'{e} approximant function having the same number of poles
as the function to be fit.  The exact number of poles, when it is not known,
can be determined from the crossing point in a $\Lambda$ vs.\ $r$ plot.
The crossing also indicates that a sufficient level of numerical accuracy
in the input points has been acheived.

There are several caveats to the procedure we have outlined.
(1) If the true Green's function has a branch cut along the real axis
arising from trancendental functions then no $\Lambda$ crossing will ever 
be observed, since a branch cut of that kind can only be represented by 
an infinity of poles ($r_0=\infty$).
(2) The self-energy of the Green's function we are trying to reproduce
must have the correct asymptotic form and must be analytic in, say,
the upper half of the complex plane; otherwise, the 
rational polynomial (or continued fraction) form of the approximant
cannot reproduce its analytic structure.
(3) The Pad\'{e} method is often used to model a function that is smooth
in some region of interest (well away from its poles) and such
calculations are rarely performed with more than machine accuracy.
Our numerical analysis of the Pad\'{e} inversion, with its prediction of
extremely high accuracy requirements, is not meant to invalidate these results.
We have applied the Pad\'{e} method to the particularly
difficult problem of reproducing the sharp peak structures characteristic 
of a spectral function whose Green's function has its poles along the 
real axis.  In that case, the poles lie in the region of interest.
The precision and accuracy requirements of the Pad\'{e} inversion are 
greatly reduced if the poles of the Green's function lie deep in the 
complex plane. 

Finally, let us remember that the starting point for our new Pad\'{e} approach
was the realization that the convergence of the continued fraction coefficients
to `allowed' values can provide a criterion for judging the quality
of a Pad\'{e} approximant, even if the analytic structure of the function
we are trying to fit is unknown.  In Sect.~IV, 
we demonstrated the utility of this idea using the $\lambda_1$ coefficient.
However, we know that there is much addtional information that can be
extracted from the remaining continued fraction coefficients.
In future, perhaps our analysis can be extended to include
$e_1$, $\lambda_2$, $e_2$, etc.

\vspace {2.0 truecm}
\acknowledgments

One of us (RJG) wishes to thank George Baker for directing him
to a variety of valuable references, and we thank David S\'{e}n\'{e}chal
and Andr\'{e}-Marie Tremblay for helpful comments on their
recent paper\cite {usherb pades}. This work was supported
by the NSERC of Canada, and the Institute of Theoretical Physics
of the University of Alberta.

\newpage

\appendix
\section*{}

In addition to the usual occupation functions
\begin{mathletters}
\begin{equation}
f[x] = \frac{1}{\beta}\sum_{\omega_n}\frac{e^{i\omega_n 0^+}}{i\omega_n-x}
= \frac{1}{e^{\beta x}+1}
\end{equation}
\begin{equation}
b[x] = -\frac{1}{\beta}\sum_{\nu_n}\frac{e^{i\nu_n 0^+}}{i\nu_n-x}
= \frac{1}{e^{\beta x}-1}
\end{equation}
\end{mathletters}
it is often convenient to define {\it partial} occupation functions.
For example, the bose version of such a function looks like
\begin{equation} \label{bose-POF}
{\tilde b}[x] = -\frac{1}{\beta}\sum_{\nu_n>\nu_p}
\frac{e^{i\nu_n0^+}}{i\nu_n-x}
= \frac{1}{2\pi i}\psi\biggl(\frac{\beta}{2\pi i}(i\nu_{p+1}-x)\biggr)
\end{equation}
where $\psi(z)=d\ln \Gamma(z)/dz$ is the digamma function\cite{Ahlfors}.
This can be generalized to a $m$-order function (symmetric
in its arguments)
\begin{equation}
{\tilde b}[x_1,x_2,\ldots,x_m] = -\frac{1}{\beta}\sum_{\nu_n>\nu_p}
\frac{1}{i\nu_n-x_1}\frac{1}{i\nu_n-x_2}\cdots\frac{1}{i\nu_n-x_m}
\end{equation}
which has the interesting property that it can be expressed (via
partial fraction decomposition) in terms of the ($m-1$)-order 
partial occupation function
\begin{equation} \label{recursive-def}
{\tilde b}[x_1,x_2,\ldots,x_m]
= \left\{\begin{array}{ll} 
\frac{{\tilde b}[x_1,x_2,\ldots,x_{m-2},x_{m-1}]-
{\tilde b}[x_1,x_2,\ldots,x_{m-2},x_m]}
{x_{m-1}-x_m} & \text{if $x_{m-1}\neq x_m$}\\
\frac{\partial}{\partial y}\bigl. 
{\tilde b}[x_1,x_2,\ldots,x_{m-2},y]
\bigr|_{y=x_m} & \text{otherwise.}
\end{array}\right.
\end{equation}
Equation (\ref{bose-POF}) serves to terminate the recursion.

Furthermore, it is straightforward to show that for all $l\ge 0$
\begin{eqnarray} \nonumber
\Theta_{(l+2)}[x] & \equiv & -\frac{1}{\beta}\sum_{|\nu_n|>\nu_p}
\frac{1}{(i\nu_n)^{l+1}}\frac{1}{i\nu_n-x}\\ \nonumber
& = & -\frac{1}{\beta}\sum_{\nu_n>\nu_p}\Biggl(
\frac{1}{(i\nu_n)^{l+1}}\frac{1}{i\nu_n-x}
+ \frac{1}{(-i\nu_n)^{l+1}}\frac{1}{-i\nu_n-x}\Biggr)\\ \nonumber
& = & {\tilde b}[\underbrace{0,0,\ldots,0}_{l+1},x]
+(-1)^l {\tilde b}[\underbrace{0,0,\ldots,0}_{l+1},-x]\\
& = & \frac{1}{l!}\frac{\partial^l}{\partial y^l}\Bigl.\Bigl\{
{\tilde b}[x,y]+(-1)^l {\tilde b}[-x,y]\Bigr\}\Bigr|_{y=0}
\end{eqnarray}
where, according to Eq.~(\ref{recursive-def}), the two-argument function 
${\tilde b}[x,y]$ is related
to ${\tilde b}[x]$ by
\begin{equation}
{\tilde b}[x,y] = \frac{{\tilde b}[x]-{\tilde b}[y]}{x-y}
\end{equation}
provided $x\neq y$.  The $\Theta$ functions provide a closed-form 
representation of the high-frequency asymptotics of a broad class of 
Matsubara sums. In particular, the sum
\begin{equation}
-\frac{1}{\beta}\sum_{\nu_{n'}}\chi^0(\vec{Q},\nu_{n'})
G^0(\vec{Q}-\vec{k},\nu_{n'}-\omega_n)
\end{equation}
can be separated into a finite sum over all low frequencies
\begin{equation}
\frac{1}{\beta}\sum_{|\nu_{n'}|\le \nu_p}\chi^0(\vec{Q},\nu_{n'})
G^0(\vec{Q}-\vec{k},\nu_{n'}-\omega_n)
\end{equation}
and an infinite sum over the remaining frequencies
\begin{eqnarray}\nonumber
-\frac{1}{\beta}\sum_{|\nu_{n'}|>\nu_p}
\chi^0(\vec{Q},\nu_{n'})G^0(\vec{Q}-\vec{k},\nu_{n'}-\omega_n)
& = & -\frac{1}{\beta}\sum_{|\nu_{n'}|>\nu_p}
\Biggl(\sum_{l=1}^{\infty}\frac{\chi^0_{(l)}(\vec{Q})}{(i\nu_{n'})^l}
\Biggr)\frac{1}{i(\nu_{n'}-\omega_n)-\xi_{\vec{Q}-\vec{k}}}\\ \nonumber
& = & -\sum_{l=1}^{\infty} \chi^0_{(l)}(\vec{Q})
\frac{1}{\beta}\sum_{|\nu_{n'}|>\nu_p} \frac{1}{(i\nu_{n'})^l}
\frac{1}{i\nu_{n'}-(i\omega_n+\xi_{\vec{Q}-\vec{k}})}\\ \label{Theta-at-work}
& = & +\sum_{l=1}^{\infty}\chi^0_{(l)}(\vec{Q})
\Theta_{(l+1)}[i\omega_n+\xi_{\vec{Q}-\vec{k}}]
\end{eqnarray}
where, in Eq.~(\ref{Theta-at-work}), we have used the fact that 
the free susceptibility $\chi^0$ admits a Laurent expansion in the
frequency variable
\begin{eqnarray} \nonumber
\chi^0(\vec{Q},\nu_n) & = & \frac{1}{i\nu_n}\frac{1}{M}\sum_{\vec{k}}
\frac{f[\xi_{\vec{k}}]+f[\xi_{\vec{Q}-\vec{k}}]-1}{1-(\xi_{\vec{k}}
+\xi_{\vec{Q}-\vec{k}})/i\nu_n}\\
& = & \frac{\chi^0_{(1)}(\vec{Q})}{i\nu_n} + 
\frac{\chi^0_{(2)}(\vec{Q})}{(i\nu_n)^2}+
\frac{\chi^0_{(3)}(\vec{Q})}{(i\nu_n)^3}+ \cdots
\end{eqnarray}
with $\vec{Q}$-dependent coefficients
\begin{equation}
\chi^0_{(l)}(\vec{Q}) = \frac{1}{M}\sum_{\vec{k}}(f[\xi_{\vec{k}}]
+f[\xi_{\vec{Q}-\vec{k}}]-1) (\xi_{\vec{k}}+\xi_{\vec{Q}-\vec{k}})^{l-1} \:.
\end{equation}

\begin{figure}
\caption{For various levels of (a) random and (b) systematic error, 
characterized roughly as $10^{-E}$ (see the text for more details),
the quality of the Pad\'{e} fit as measured by $F$ 
(see Eq.~(\protect\ref{eq:F eqn})) is plotted with respect 
to the number of poles in the Pad\'{e} approximant (the solid lines
are a guide to the eye). The ${\vec k = 0}$ self energy
being studied is that of Eq.~(\protect\ref{T-matrix-sigma}) where the
parameters of the attractive Hubbard model (with $t$ being the hopping
energy), for an $8\times 8$ square
lattice, are a repulsive energy $|U|/t=4$, a chemical potential 
$\mu/t = -2$, and a temperature $T/t=0.7$. The vertical dashed line 
indicates the number of poles ($r_0=26$) in the true Green's function.
In plot (a), error bars (representing the standard deviation of the data points 
over a set of initial random seeds) are smaller than the symbols marking the 
data points and are not shown.  In plot (b), the dotted line is the best linear
fit through the maximum values of $F$.}
\end{figure}

\begin{figure}
\caption{The ${\vec k = 0}$ spectral function of the Pad\'{e} approximant is 
compared to the 
exact spectral function for different levels of random error ($10^{-E}$) on 
the initial input 
points. The parameters of the Hamiltonian and the self-energy being
studied are the same as those of Fig. 1.}
\end{figure}

\begin{figure}
\caption{For various levels of (a) random and (b) systematic error ($10^{-E}$),
the parameter $\Lambda$ is plotted with respect 
to the number of poles in the Pad\'{e} approximant.
The parameters are the same as those of Fig. 1.
The vertical dashed line 
indicates the number of poles ($r_0=26$) in
the true Green's function.  The solid line originating in the lower left corner 
is given by $2r\log10 (4r-1)\pi T$.  In plot (b), the dotted line is the best 
linear fit through the maximum values of $\Lambda$.
The parameters of the Hamiltonian and the self-energy being 
studied are the same as those of Fig. 1.}
\end{figure}


\begin{references}
\bibitem{Self-energy-note} We have chosen to absorb all frequency 
independent terms of the self-energy into $\xi$.
\bibitem{Linear-note} The system of equations is linear in the 
$(p,q)$ basis, but not in the $(\lambda,e)$ basis.
\bibitem{250} There is no reason to choose this specific value. Any
{\tt Digits} which is in excess of what is required by the statistical
test discussed in this section is adequate. It is important to note
that there is virtually no extra time required to make such
high accuracy calculations, and thus there is no reason to not implement
algorithms such as this with a very high accuracy.
\bibitem{Mat-freq-note}  We use the convention that 
$\omega_n = (2n-1)\pi/\beta$ represents a fermionic Matsubara frequency 
and $\nu_n = 2n\pi/\beta$ a bosonic one.
\bibitem{identity} These relations can be derived most easily using the
identity $(1-f(x)-f(y))*N(x+y)=f(x)*f(y)$, where $f$/$N$ is the fermi/bose
distribution function.
\bibitem{Param-choice-note} As viewed in the $\mu$--$T$ phase diagram,
this choice of parameters places the system well inside the band 
($ -4 < \mu/t < 4$) and just above the superconducting instability ($T>T_c$), 
the so-called Thouless criterion line. This region is of interest in that one 
expects to see a suppression of the spectral weight around the chemical 
potential, the so-called normal state pseudo-gap.
\bibitem{Freq-sum-note} In fact, the frequency sum can be performed formally 
to yield an infinite sum of generalized occupation functions.  However, 
this is of no computational value since it only trades a difficult frequency 
sum for an infinite series of increasingly
more difficult (at each order) $\vec{k}$ sums.
\bibitem{series} This expansion is easy to implement in MAPLE using the 
{\tt series()} command, although the computational resources required to 
execute it are significant due to the combinatorial explosion of terms at 
high order.
\bibitem{sc paper} The details of these calculations will be presented in
a separate publication.
\end{references}
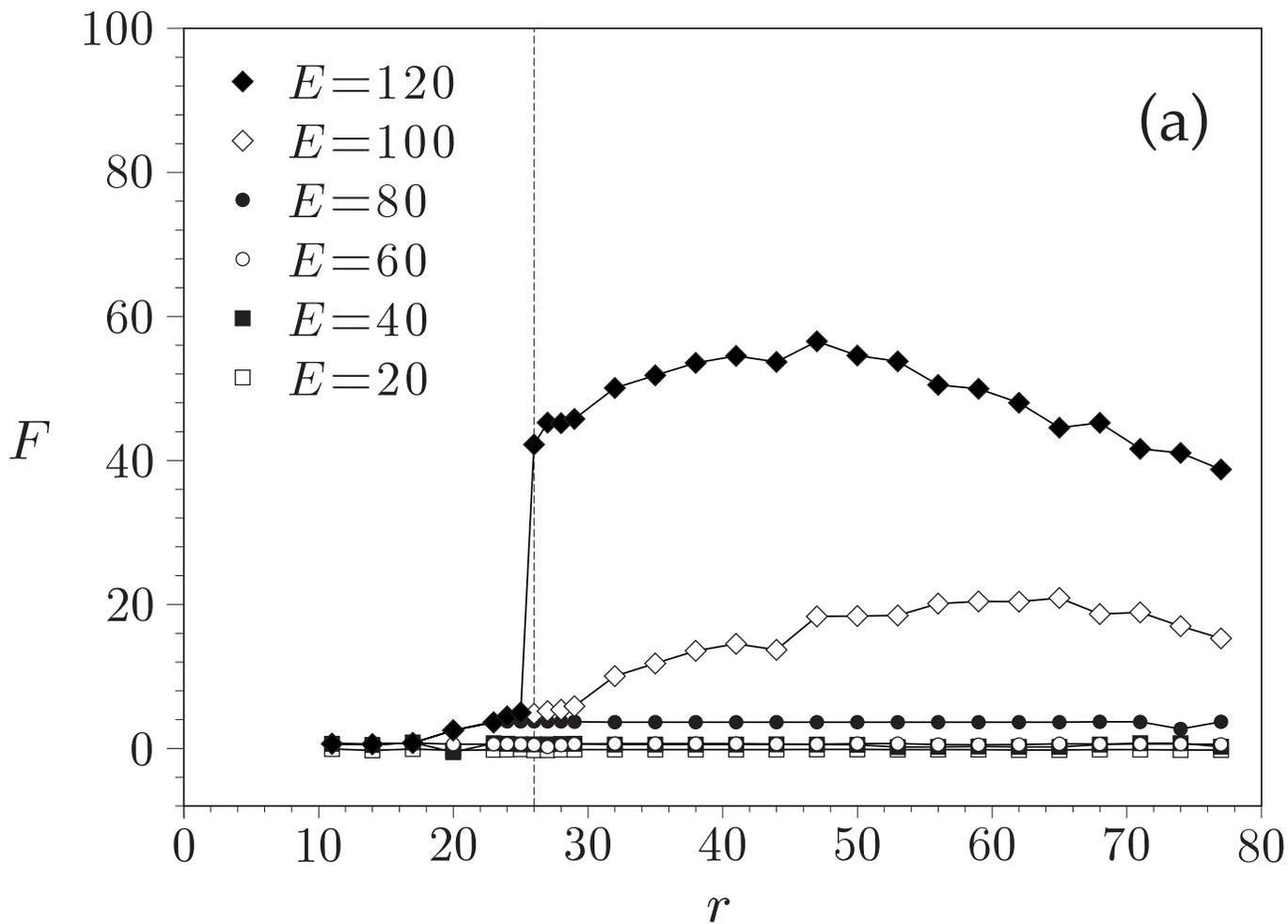
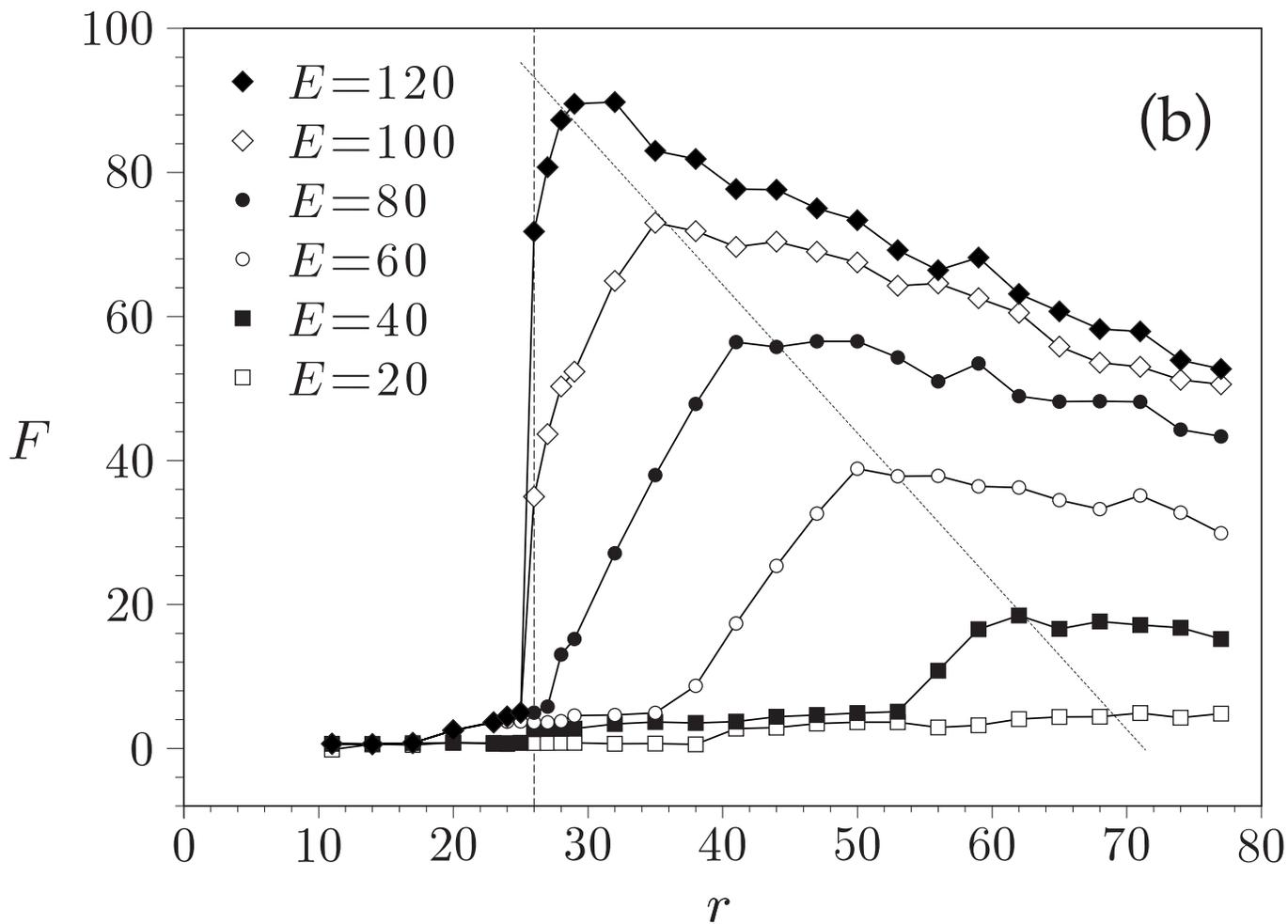
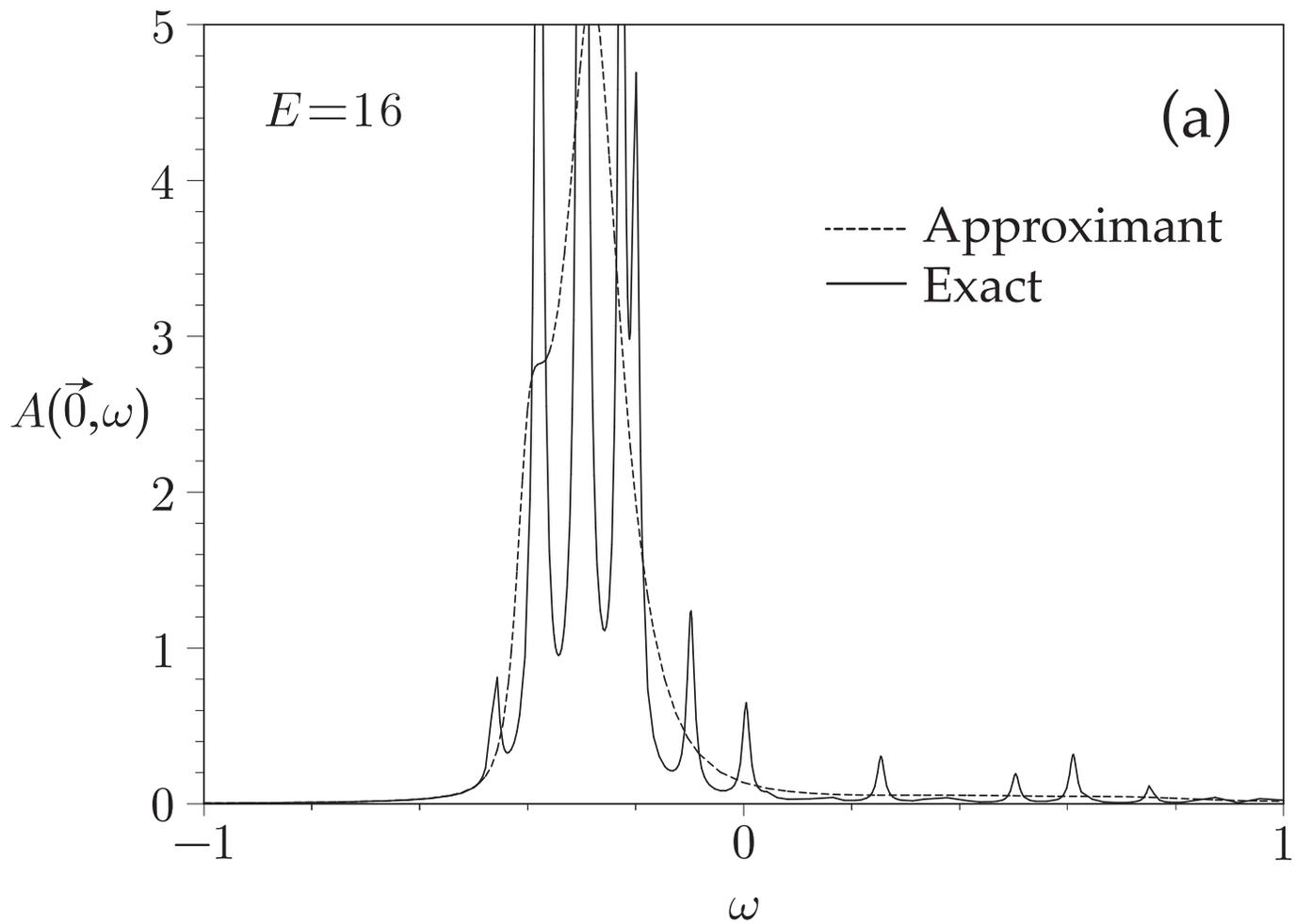
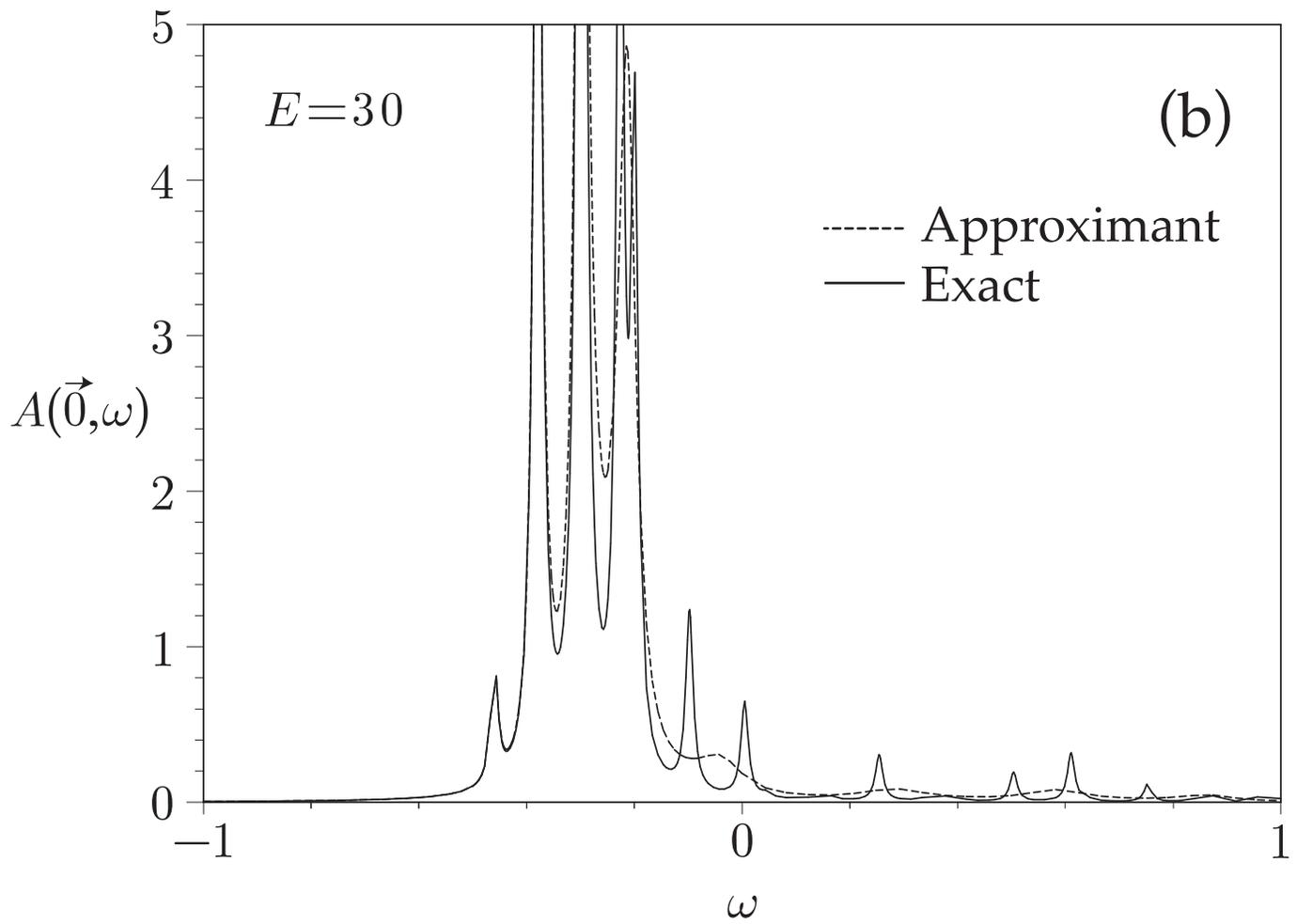
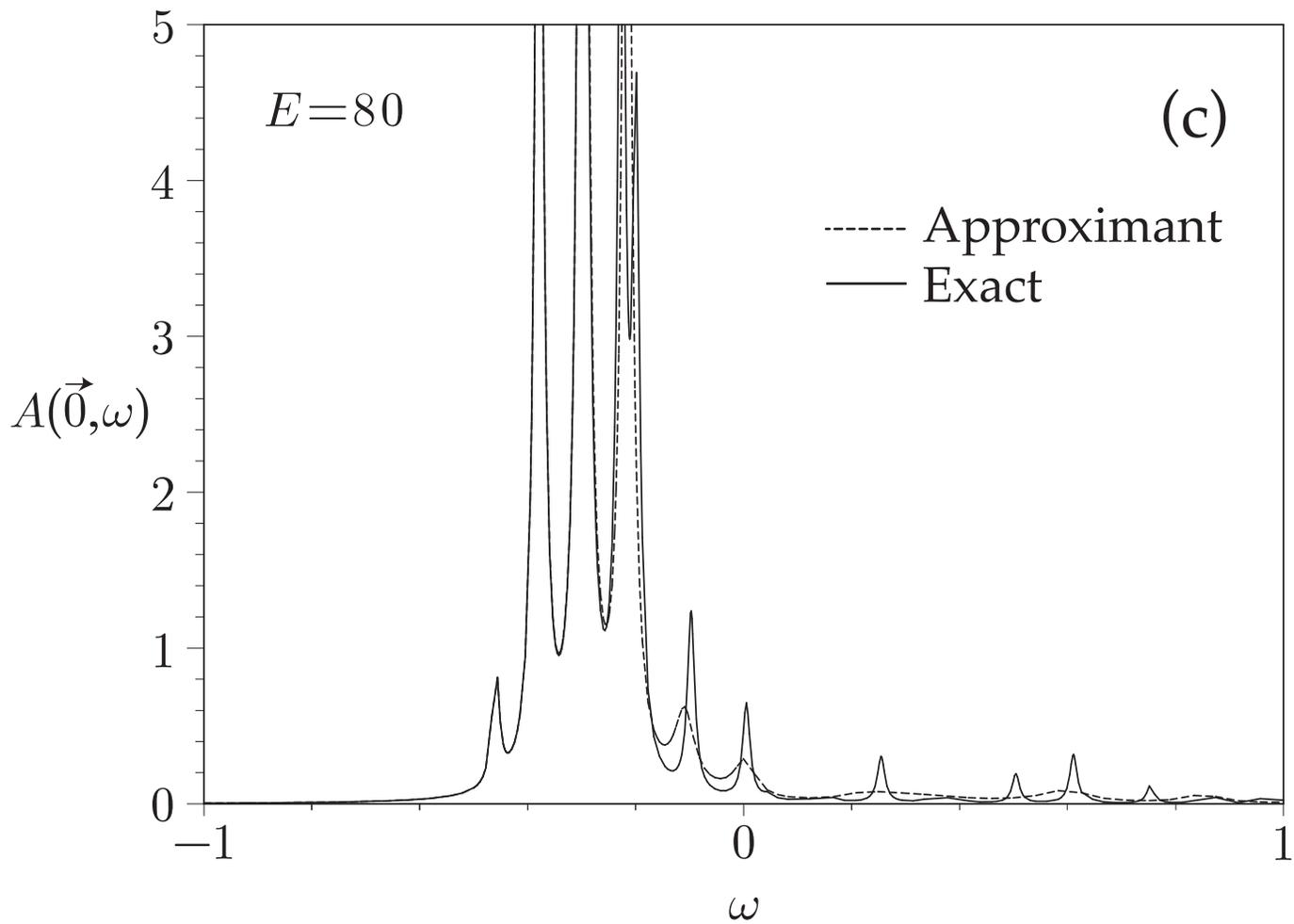
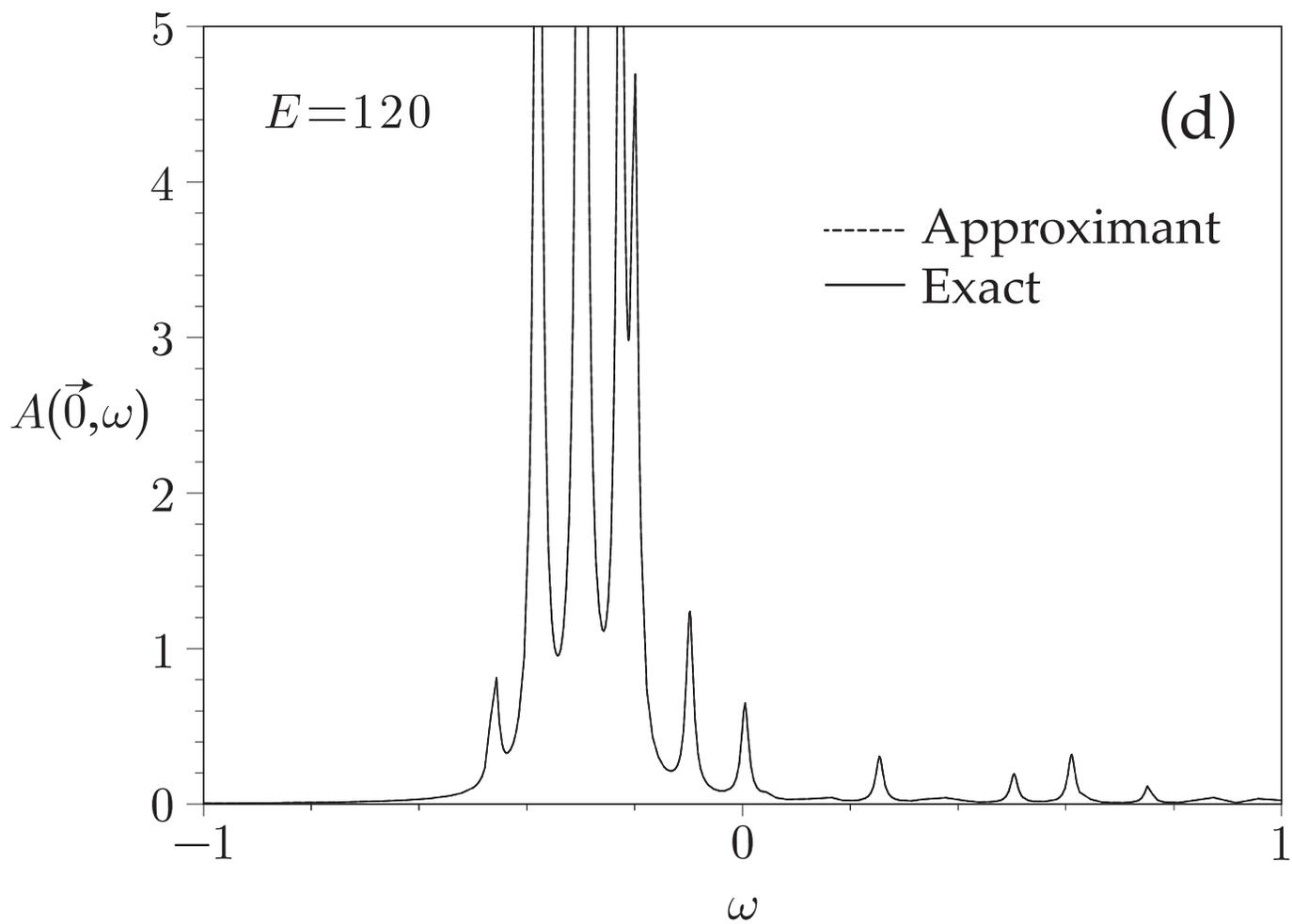
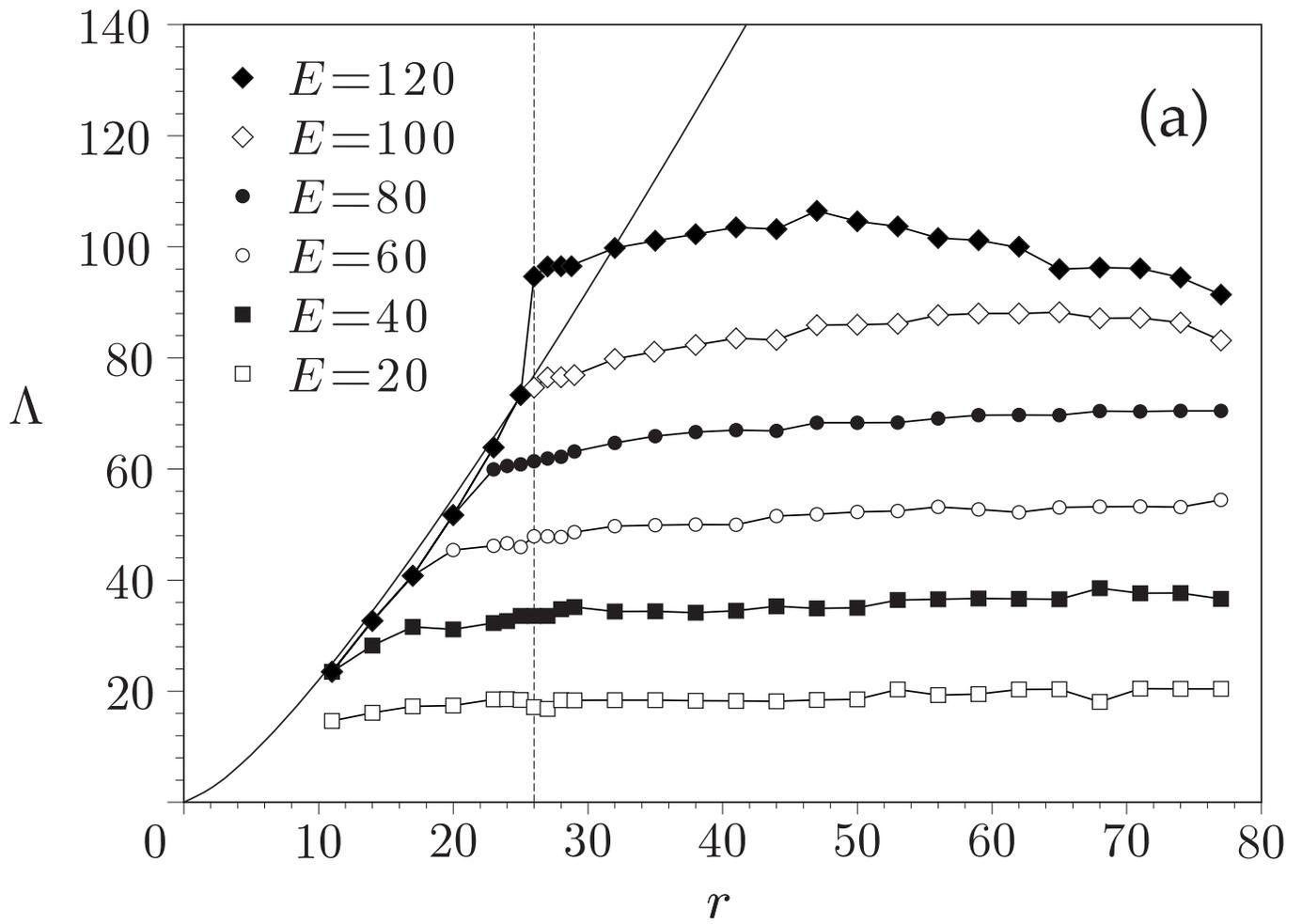
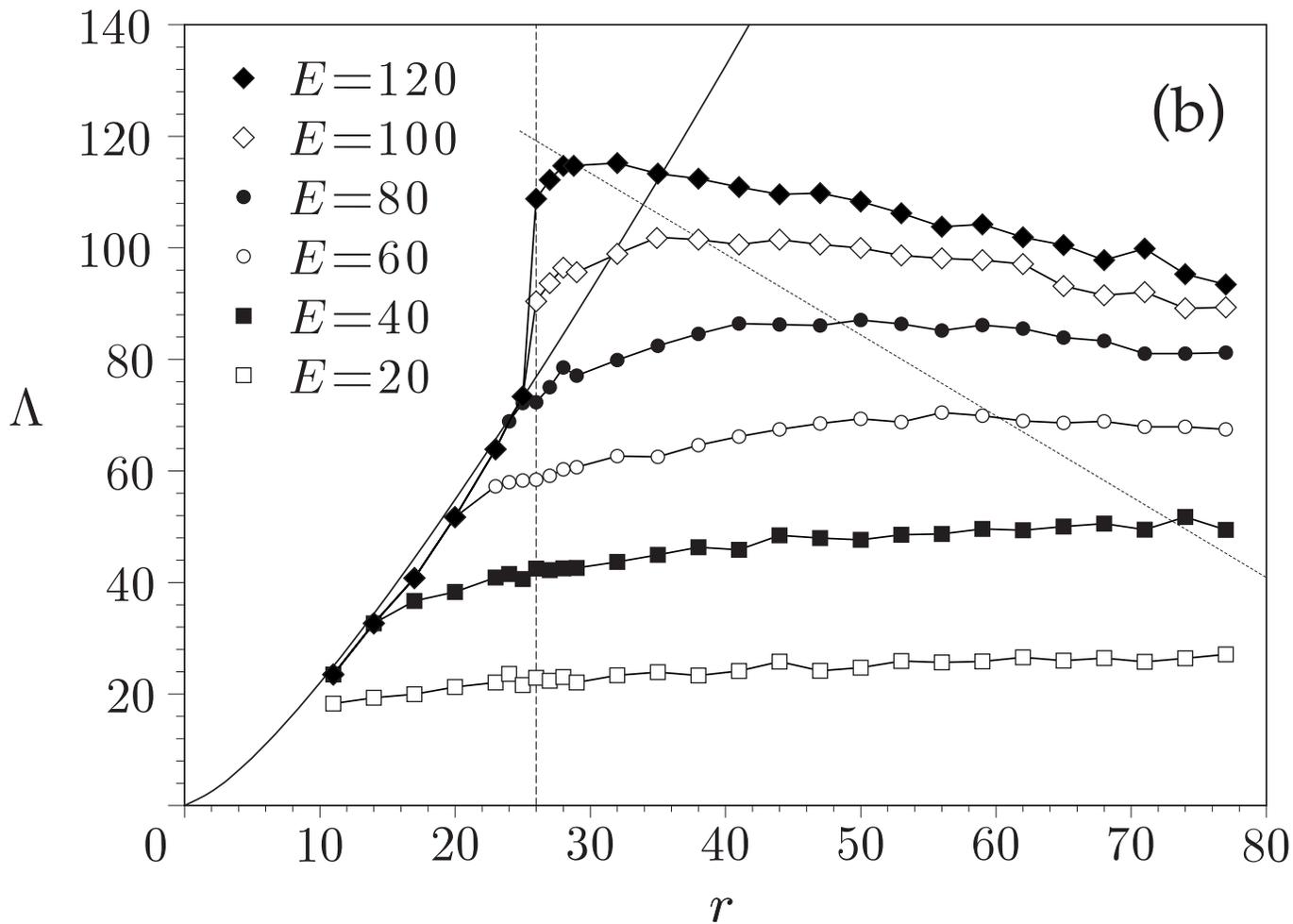
\end{document}